**Exact solutions of Klein–Gordon equation for the Makarov potential with the asymptotic iteration method**

**M. Chabab and M. Oulne**

LPHEA, Department of Physics, FSSM, Cadi Ayyad University,
P.O.B 2390, Marrakech 40 000, Morocco
E-mail: mchabab@ucam.ac.ma; oulne@ucam.ac.ma

**Abstract:** we derive exact analytical solutions of the Klein-Gordon equation for Makarov potential by means of the asymptotic iteration method. The energy eigenvalues are given in a closed form and the corresponding normalized eigenfunctions are obtained in terms of the generalized Laguerre polynomials and hypergeometrical functions .

## I. Introduction

The Klein – Gordon equation (KG) is nowadays regarded as the relativistic form of the Schrödinger equation. It affords appropriate description for spin zero particles. Since the solution of the KG equation is often a complicated problem, use of mathematical methods is required. These include the variational method [1], the functional analysis method [2], the supersymetric approach [3], the Nikiforov – Uvarov method (NU) [4], the shifted 1/N expansion [5] and the asymptotic iteration method (AIM) [6]. The latter, however, is particularly suitable for obtaining solutions to such differential equations. Indeed, compared to other analytic techniques, the AIM offers an accurate and efficient investigation of the spectrum of many particles in relativistic and non relativistic quantum mechanics [7-10].

The KG equation has been investigated for several potentials as the generalized Woods-Saxon potential [11], Hulthen potential [12], pseudo and perturbed Coulomb potential [13-14] and equal scalar and vector potential [4]. In the present paper, we extend the investigation to Makarov potential [15] using the asymptotic iteration method [16]. The Makarov potential [15] offers an appropriate description of ring shaped molecules such as benzene $C_6H_6$ and provides an interesting interaction model of deformed pairs of nuclei.

## II. Theory

In spherical coordinates, the KG equation for a particle in general non-central potential$V(r,\theta)$ can be written as

$$\left[\frac{1}{r^2}\frac{\partial}{\partial r}\left(r^2\frac{\partial}{\partial r}\right)+\frac{1}{r^2 sin\theta}\frac{\partial}{\partial\theta}\left(sin\theta\frac{\partial}{\partial\theta}\right)+\frac{1}{r^2 sin^2\theta}\frac{\partial^2}{\partial\theta^2}-(E+M)V(r,\theta)\right]\psi(r,\theta,\varphi)=0 \quad (1)$$

where $E$ is the energy and $M$ the mass of the particle. The non-central potential of Makarov [15] reads

$$V(r,\theta)=-\frac{\alpha}{r}+\frac{\beta}{r^2 sin^2\theta}+\frac{\gamma cos\theta}{r^2 sin^2\theta} \quad (2)$$

where the first term represents the Coulomb potential while the second and the third are the short range ring shape terms.

As usual we can write the wave function in the following form

$$\psi(r,\theta,\varphi)=\frac{R(r)}{r}\Theta(\theta)\Phi(\varphi) \quad (3)$$

then, the equation (1) will be separated into a radial part

$$\left[\frac{d^2}{dr^2}-\frac{\lambda}{r^2}-(E+M)\frac{\alpha}{r}+E^2-M^2\right]R(r)=0 \tag{4}$$

and angle dependent equations

$$\left[\frac{d^2}{d\theta^2}+cot\theta\frac{d}{d\theta}+\lambda-\frac{m^2}{sin^2\theta}-(E+M)\frac{\beta+\gamma cos\theta}{sin^2\theta}\right]\Theta(\theta)=0 \tag{5a}$$

$$\left[\frac{d^2}{d\varphi^2}+m^2\right]\Phi(\varphi)=0 \tag{5b}$$

where $m^2$ and λ are separation constants. The solution of equation (5b) is well-known

$$\Phi(\varphi)=Ae^{im\varphi}, \ (m=0, \pm1, \pm2, \ldots) \tag{6}$$

Only Eq. (4) and Eq.(5a) have to be solved. To this end, we use the asymptotic iteration method and the first step consists in the conversion of these equations to standard forms suitable to AIM applications [16]. So the radial equation (4) is rearranged as

$$\left[\frac{d^2}{dr^2}-\frac{\ell(\ell+1)}{r^2}+\frac{2s}{r}+k^2\right]R(r)=0 \tag{7a}$$

with

$$\lambda=\ell(\ell+1),\ s=\frac{(E+M)\alpha}{2}\quad \text{and}\quad k^2=E^2-M^2 \tag{7b}$$

and where the boundary condition is $R(0)=0$.

Now, the asymptotic behavior of the radial wave function suggests the following ansatz for $R(r)$

$$R(r)=A(kr)^{\ell+1}e^{ikr}f(r) \tag{8}$$

where $A$ is a normalization constant.

For this form of the wave function, the radial equation (7a) reads

$$\frac{d^2}{dr^2}f(r)=\lambda_0(r)\frac{d}{dr}f(r)+s_0(r)f(r) \tag{9}$$

with

$$\lambda_0(r)=-\left(\frac{2\ell+2+2ikr}{r}\right) \tag{10a}$$

$$s_0(r)=-\left(\frac{2ikr\,(\ell+1)+2s}{r}\right) \tag{10b}$$

If, moreover, we introduce the new variable$x=-2ikr$, these equations transform into

$$\frac{d^2}{dx^2}f(x) = \lambda_0(x)\frac{d}{dx}f(x) + s_0(x)f(x) \tag{11}$$

$$\lambda_0(x) = -\left(\frac{2\ell+2-x}{x}\right) \tag{12a}$$

$$s_0(x) = \left(\frac{\ell+1-\frac{is}{k}}{x}\right) \tag{12b}$$

According to the AIM procedure, the energy eigenvalues are then computed by means of the following termination condition [16]

$$\delta = s_n\lambda_{n-1} - \lambda_n s_{n-1} = 0 \tag{13}$$

for a given x > 0, with the sequences

$$\lambda_n(x) = \lambda_{n-1}^{'}(x) + s_{n-1}(x) + \lambda_0(x)\lambda_{n-1}(x) \tag{14a}$$

$$s_n(x) = s_{n-1}^{'}(x) + s_0(x)\lambda_{n-1}(x), \text{ n= 1, 2, 3, …} \tag{14b}$$

For n = 10 iterations, the obtained solutions are

$$k = \frac{is}{\ell+1}, k = \frac{is}{\ell+2}, k = \frac{is}{\ell+3}, k = \frac{is}{\ell+4}, \ldots\ldots. \tag{15}$$

In general form, we have

$$k = \frac{is}{\ell+1+N}, N = 0, 1, 2, \ldots.. \tag{16}$$

Substituting k and s by their expressions given in Eq. (7b), we finally derive the exact eingenvalues $E_N$ of the radial part of the Klein Gordon equation

$$E_N = M\frac{(\ell+1+N)^2-\frac{\alpha^2}{4}}{(\ell+1+N)^2+\frac{\alpha^2}{4}} \tag{17}$$

where N is the radial quantum number.

Now we turn solutions of the polar angle equation (5a). By introducing a new variable $y = \frac{1}{2}(1+cos\theta)$, it becomes

$$\left[y(1-y)\frac{d}{dy}y(1-y)\frac{d}{dy} + \ell(\ell+1)y(1-y) - a^2 - (b^2-a^2)y\right]G(y) = 0 \tag{18}$$

with

$$a = \frac{\sqrt{m^2+\beta'^2-\gamma'^2}}{2}, \quad b = \frac{\sqrt{m^2+\beta'^2+\gamma'^2}}{2} \tag{19}$$

$\beta^{'} = (E+M)\beta$ and $\gamma^{'} = (E+M)\gamma$. Since G(y) must satisfy the boundary conditions G(0) = G(1) = 0, we use the following ansatz

$$G(y) = By^a(1-y)^b f(y) \tag{20}$$

Substitution of this wave function into Eq. (18) leads to the AIM standard forms

$$\frac{d^2}{dy^2} f(y) = \lambda_0(y)\frac{d}{dy} f(y) + s_0(y) f(y) \tag{21}$$

$$\lambda_0(y) = -\frac{2a+1-2(a+b+1)}{y(1-y)} \tag{22a}$$

$$s_0(y) = \frac{(a+b)(a+b+1)-\ell(\ell+1)}{y(1-y)} \tag{22b}$$

We may then derive the eigenvalues of Eq. (18) by considering AIM equation (12) for a given y > 1. For 10 iterations, we obtain

$$\ell = a + b, \ell = a + b + 1, \ell = a + b + 2, \ell = a + b + 3, \dots \tag{22}$$

Or equivatently

$$\ell = a + b + n, n = 0, 1, 2, 3, \dots.. \tag{23}$$

Substituting a and b by their expressions given in (19) leads to

$$\ell_n = \sqrt{\frac{m^2+\beta'+\sqrt{(m^2+\beta')^2-\gamma'^2}}{2}} + n \tag{24}$$

Once this condition for $\ell$ is inserted in the previously derived expression (17) we get the exact energy eigenvalues for a bound particle in the Makarov potential

$$E_{Nnm} = M\frac{\left(\sqrt{\frac{m^2+\beta'+\sqrt{(m^2+\beta')^2-\gamma'^2}}{2}}+n+1+N\right)^2-\frac{\alpha^2}{4}}{\left(\sqrt{\frac{m^2+\beta'+\sqrt{(m^2+\beta')^2-\gamma'^2}}{2}}+n+1+N\right)^2+\frac{\alpha^2}{4}} \tag{25}$$

The obtained energy spectrum is identical to the one determined in the paper [4] via the Nikiforov - Uvanov method.

Next, we proceed to look for the radial wave function *R(r)*. For this, we have to determine the function *f(x)* by solving the differential equation (11). Substituting k by its expression given in Eq. (16), this equation transforms into Kummer equation

$$\frac{d^2}{dx^2} f(x) + \left(\frac{2\ell+2-x}{x}\right)\frac{d}{dx} f(x) + Nf(x) = 0 \tag{26}$$

whose solutions as $x \to 0$ are the confluent hypergeometrical functions,

$$f(x) = C_N F(-N, 2\ell + 2, x) \tag{27}$$

Thanks to the relation between Kummer functions F (a, b; x) and the generalized Laguerre polynomials $L_N^{2\ell+1}(x)$, the radial wave function can be written

$$R_{N\ell}(x) = D_{N\ell}x^{\ell+1}e^{-\frac{x}{2}}L_N^{2\ell+1}(x) \tag{28}$$

where $D_{N\ell}$ is a normalization constant computed via the orthogonality relation of Laguerre polynomials [17]

$$D_{n'\ell} = \left[\frac{(E+M)\alpha(n'-\ell-1)!}{(n')^2\Gamma(n'+\ell+1)}\right]^{1/2} \tag{29}$$

where $n' = N + \ell + 1$ and finally we derive the expression of the radial wave function

$$R_{n'\ell}(r) = \left[\frac{(E+M)\alpha(n'-\ell-1)!}{(n')^2\Gamma(n'+\ell+1)}\right]^{1/2} \left(\frac{(E+M)\alpha}{n'}\right)^{\ell+1} r^{\ell+1}e^{-\frac{(E+M)\alpha}{2n'}r} L_{n'-\ell-1}^{2\ell+1}\left(\frac{(E+M)\alpha}{n'}r\right) \tag{30}$$

Similarly, the angular wave function G(y) is obtained by solving the hypergeometric differential equation whose solutions are given by [17]

$$f(y) = {}_2F_1(-n, a+b+\ell+1; 2a+1; y) \tag{31}$$

where ${}_2F_1$ is a special case of the generalized hypergeometric function. Therefore, by combining Eqs. (20) and (31), the polar angle wave functions are found to be

$$G_{\ell m}(y) = N_{\ell m}y^a(1-y)^b \ {}_2F_1(-n, a+b+\ell+1; 2a+1; y) \tag{32}$$

From the orthonormality of hypergeometric functions, we determined the normalization constant

$$N_{\ell m} = \frac{1}{\Gamma(2a+1)}\left[\frac{\Gamma(\ell+a+b+1)\Gamma(\ell+a-b+1)(2\ell+1)}{2(\ell-a-b)!\Gamma(\ell-a+b+1)}\right]^{1/2} \tag{33}$$

where the parameters a and b are provided by Eq.(19).

## III. Conclusion

In the present paper, the bound states of the Klein Gordon equation for Makarov potential are investigated. Using the asymptotic iteration method, we have derived the exact expression of the energy eigenvalues and the corresponding normalized eigenfunctions in terms of the generalized Laguerre polynomials and hypergeometrical functions. The obtained analytical results are very useful in nuclear physics and quantum chemistry.

**Acknowledgments:**

This study is partially supported by the CNRST Moroccan research program PROTAS III, D16/04.